\documentclass{ws-mpla}

\begin{document}

\renewcommand{\draftnote}{} 
\renewcommand{\trimmarks}{} 

\markboth{A. M. Green}
{Astrophysical uncertainties on direct detection experiments}

\catchline{}{}{}{}{}

\title{Astrophysical uncertainties on direct detection experiments}

\author{\footnotesize Anne M. Green}

\address{School of Physics and Astronomy, University of Nottingham,
  University Park, Nottingham, NG7 2RD, UK\\
anne.green@nottingham.ac.uk}

\maketitle


\begin{abstract}
  Direct detection experiments are poised to detect dark matter in the
  form of weakly interacting massive particles (WIMPs). The signals
  expected in these experiments depend on the ultra-local WIMP density
  and velocity distribution. Firstly we review methods for modelling the dark
  matter distribution.  We then discuss observational determinations
  of the local dark matter
  density, circular speed and escape
  speed and the results of numerical simulations of Milky Way-like
  dark matter halos. In each case we highlight the uncertainties and
  assumptions made. We then overview the resulting uncertainties in
  the signals expected in direct detection experiments, specifically
  the energy, time and direction dependence of the event rate.
  Finally we conclude by discussing techniques for handling
  the astrophysical uncertainties when interpreting data from direct detection experiments.
\end{abstract}

\ccode{95.35.+d,98.35.Gi,}

\section{Introduction}	

Weakly Interacting Massive Particles (WIMPs) are a promising dark
matter candidate as they are generically produced in the early
Universe with roughly the right density. Furthermore supersymmetry (SUSY) provides a
well-motivated concrete WIMP candidate in the form of the lightest
neutralino (for reviews see e.g. Refs.~\refcite{jkg,bhs,dmbook}). WIMPs
can be directly detected in the lab via their elastic scattering off
target nuclei~\cite{dd}. Numerous direct detection experiments are
underway and they are probing the regions of WIMP mass-cross-section
parameter space populated by SUSY models (see
e.g. Ref.~\refcite{theory}).

The realisation that uncertainties in the velocity distribution,
$f({\bf v})$, will affect the direct
detection signals dates back to some of the earliest direct detection
papers written in the 1980s (e.g. Ref.~\refcite{dfs}). We first
discuss the standard halo model and other approaches to modelling the
Milky Way halo (Sec.~\ref{modelling}). We then discuss what
observations and simulations tell us about the dark matter distribution (Secs.~\ref{astro1} and \ref{astro2}
respectively). We then examine the direct detection signals (Sec.~\ref{signals})
and how the astrophysical uncertainties affect these
signals (Sec.~\ref{affect}). We conclude in Sec.~\ref{strategies} by
discussing strategies for handling the uncertainties, including both
`astrophysics independent' approaches and parameterizing the WIMP
speed distribution.

\section{Halo modelling}
\label{modelling}

\subsection{Standard halo model}
\label{shm}


The steady-state phase space distribution function, $f$, of a collection of
collisionless particles is given by the solution of the collisionless
Boltzmann equation:
\begin{equation}
\frac{{\rm d} f}{{\rm d} t}  = 0 \,.
\end{equation}
In Cartesian coordinates this becomes 
\begin{equation}
\label{cme}
\frac{\partial f}{\partial t} + {\bf
  v}.  \frac{\partial f}{\partial {\bf x}} - \frac{\partial
  \Phi}{\partial {\bf x}} \frac{\partial f}{\partial {\bf v}} = 0 \,,
\end{equation}
where $\Phi$ is the potential.

The standard halo model (SHM) is an isotropic, isothermal sphere with
density profile $\rho(r) \propto r^{-2}$.  In
this case the solution of the collisionless Boltzmann equation is a
so-called Maxwellian velocity distribution, given by 
\begin{equation}
\label{max}
f({\bf v}) = N
 \exp{\left( - \frac{3|{\bf v}|^2}{2 \sigma^2} \right)}  \,, \\
\end{equation}
where $N$ is a normalisation constant. The
isothermal sphere has a flat rotation curve at large
radii and the velocity dispersion is related to the asymptotic value
of the circular speed (the speed with which objects on circular orbits
orbit the Galactic centre)  $v_{{\rm c}, \infty}= \sqrt{2/3}  \, \sigma$.  It
is usually assumed that the rotation curve has already reached its
asymptotic value at the Solar radius, $r=R_{0}$, so that $\sigma=
\sqrt{3/2} \, v_{\rm c}$ where $v_{\rm c}\equiv v_{\rm c}(R_{0})$ is the
local circular speed.  In the SHM the peak speed $v_{0}$ and the
circular speed are identical, $v_{\rm c}=v_{0}$, and these parameters
are often used interchangeably. However this is not the case in
general, for instance for a NFW~\cite{nfw} density profile $v_{\rm c} =
0.88 v_{0}$ at $r=r_{\rm s}/2$ (where the scale radius, $r_{\rm s}$,
is the radius at which the logarithmic slope of the density profile is equal to -2)~\cite{kuhlen}.

The density distribution of the SHM is formally infinite and hence the
velocity distribution extends to infinity too. In reality the Milky
Way halo is finite, and particles with speeds greater than the escape
speed, $ v_{\rm esc}(r) = \sqrt{2 |\Phi(r)|}$, will not be gravitationally bound to the
MW. This is often addressed by simply truncating the velocity distribution at
the measured local escape speed, $v_{\rm esc} \equiv v_{\rm esc}(R_{0})$, so that $f({\bf v})$ is
given by eq.~(\ref{max}) for $|{\bf v}| < v_{\rm esc}$ and $f({\bf
  v}) = 0$ for $|{\bf v}| \geq v_{\rm esc}$. This sharp truncation is
clearly unphysical. An alternative, but still ad hoc, approach is to
make the truncation smooth:
\begin{equation}
\label{max2}
f({\bf v}) =
\begin{cases}
 N \left[
 \exp{\left( - \frac{3|{\bf v}|^2}{2 \sigma^2} \right)}  \exp{\left( -
     \frac{3 v_{\rm esc}^2}{2 \sigma^2} \right)}   \right] \,,  &
|{\bf v}| < v_{\rm esc} \,, \\
 0 \,, &  |{\bf v}| \geq v_{\rm esc} \,.
\end{cases}
\end{equation}
Another approach, used in the King model/lowered
isothermal sphere, is to modify the SHM distribution function so that
it becomes zero for large energies~\cite{bt}.

The standard parameter values used for the SHM are a local density
$\rho_{0} \equiv \rho(R_{0}) = 0.3 \, {\rm GeV} \, {\rm cm}^{-3}$, a local circular
speed $v_{\rm c} = 220 \, {\rm km \, s}^{-1}$~\cite{kerr}, and a local
escape speed $v_{\rm esc}=544 \, {\rm km \, s}^{-1}$~\cite{smith}.  We
will discuss the determination of these parameters and the associated
uncertainties in
Sec.~\ref{astro1}.

\subsection{Non-standard halo models}

For spherical, isotropic systems there is a unique relationship
between the density profile and the distribution function 
known as the Eddington equation (see
e.g. Ref.~\refcite{bt,uk}):
\begin{equation}
f(\epsilon) = \frac{1}{\sqrt{8} \pi^2} \left[ \int_{0}^{\epsilon} \frac{{\rm
      d} \Psi}{\sqrt{\epsilon - \Psi}} \frac{{\rm d}^2 \rho}{{\rm d}
    \Psi^2} + \frac{1}{\sqrt{\epsilon}} \left( \frac{{\rm d} \rho}{{\rm d}
      \Psi} \right)_{\Psi=0} \right]\,,
\end{equation}
where $\Psi(r)= - \Phi(r)+\Phi(r=\infty)$ and $\epsilon= -
E_{\rm kin} + \Psi(r)$, where $E_{\rm kin}$ is the kinetic energy.
The Eddington equation has been used to calculate the speed distribution for a
range of spherically symmetric density profiles~\cite{uk,vo1,vo2,cu2}.
Ref.~\refcite{cu2} found, using a Bayesian analysis incorporating
various dynamical constraints, speed distributions that are not too
dissimilar to the standard Maxwellian (when the same value of the
circular speed is used).  Ref.~\refcite{lsww} used the Eddington
equation to motivate a phenomenological form for $f(|{\bf v}|)$ which matches
the high speed tail of the speed distributions found in simulated halos
(see Sec.~\ref{astro2}).

The Osipkov-Merritt models~\cite{om} assume that the distribution
function depends on the energy, $\epsilon$, and angular momentum, $L$, only via a single
parameter, $Q$,
\begin{equation}
Q = \epsilon - \frac{L^2}{2 r_{a}^2} \,,
\end{equation}
where the constant $r_{a}$ is the anisotropy radius.
In this model the velocity anisotropy
\begin{equation}
\label{beta}
\beta = 1 - \frac{\sigma_{\theta}^2 + \sigma_{\phi}^2}{2
\sigma_{r}^2} \,,
\end{equation}
where $\sigma_{\theta, \phi, r}$ are the velocity dispersions in the
$r, \theta$ and $\phi$ directions,
has a particular, radially dependent, form:
\begin{equation}
\beta = \frac{1}{1 + (r_{a}^2/r^2)} \,,
\end{equation}
and the distribution function can be
calculated from a modified version of the Eddington equation
(see e.g. Refs.~\refcite{bt,uk}). In these models the peak of the
speed distribution
becomes narrower and there is an excess of high speed particles (e.g. Ref.~\refcite{greenexclude}).

In general there is no unique relationship between the density profile
and the velocity distribution for triaxial and/or anisotropic systems. In these cases a common
approach is to use the Jeans equations to calculate the lower order
moments of the distribution function. Multiplying the collisionless
Boltzmann equation, eq.~(\ref{cme}), by one of the velocity components
and integrating gives the Jeans equations, which in Cartesian
coordinates take the form
\begin{equation}
\frac{\partial (\rho \overline{v}_{j})}{\partial t} + \frac{\partial (\rho
  \overline{v_{j} v_{i}} )}{\partial x_{i}} + \rho  \frac{\partial \Phi}{\partial x_{j}} =0\,.
\end{equation}
We now have three equations (corresponding to $j=1,2,3$) for six
unknowns ($\overline{v_{1}^{2}}, \overline{v_{2}^{2}}, \overline{v_{3}^{2}},
\overline{v_{1}v_{2}}, \overline{v_{2} v_{3}}, \overline{v_{1}v_{3}}$). To make
further progress it is therefore necessary to make assumptions about
the alignment of the velocity ellipsoid, i.e. to choose coordinates
such that $\overline{v_{i} v_{j}}=0$ for $i \neq j$. The velocity
distribution is then approximated by a multivariate gaussian in these
coordinates:
\begin{equation}
f({\bf v}) \propto \exp{ \left( - \frac{v_{1}^2}{2 \sigma_{1}^2}  -
    \frac{v_{2}^2}{2 \sigma_{2}^2}   - \frac{v_{3}^2}{2 \sigma_{3}^2}
  \right)} \,, 
\end{equation}
where $\sigma_{i}^2 \equiv \overline{v_{i}^{2}}$.

Evans, Carollo and de Zeeuw~\cite{ecz} studied the logarithmic
ellipsoidal model, the simplest triaxial generalisation of the
isothermal sphere.  They argue that aligning the velocity ellipsoid
with conical coordinates is physically well motivated. In the planes
of the halo, conical coordinates are locally equivalent to cylindrical
polar coordinates. Hence, in this model, the velocity distribution can
be approximated by a multi-variate gaussian in cylindrical polar
coordinates, with velocity dispersions depending on the shape of the
halo, the velocity anisotropy and also location within the halo. Both
the width of the speed distribution and the peak speed can change
(e.g. Refs.~\refcite{ecz,greenexclude}).

Analytic models inevitably make assumptions (regarding the shape and
anisotropy of the halo, and their radial dependence) which are almost
certainly not completely valid. The relationship between the observed
properties of dark matter halos and the velocity distribution is non-trivial; models with
the same bulk properties (e.g. shape and local velocity anisotropy)
can have velocity distributions with very different forms.
Furthermore all analytic approaches to halo modelling rely on the
assumption that the phase space distribution function has reached a
steady state. To some extent analytic halo models have been superseded
by results from numerical simulations. It should be emphasized (see
Sec.~\ref{micro} and \ref{baryon}) that velocity distributions from
numerical simulations also involve approximations and extrapolations.

\section{Observations}
\label{astro1}

\subsection{Density}

The event rate is directly proportional to the local density,
$\rho_{0}$.  The standard value used is $\rho_{0}
= 0.3 \, {\rm GeV} \, {\rm cm}^{-3} = 0.008 M_{\odot} \, {\rm pc}^{-3} =
5 \times 10^{-25} \, {\rm g} \, {\rm cm}^{-3}$. As discussed in
Ref.~\refcite{salucci}, the origin of the use
of this particular value, rather than say $\rho_{0} = 0.4 \, {\rm GeV}
\, {\rm cm}^{-3}$, is unclear.

The local density is calculated via mass modelling of the Milky Way
(MW).  This involves constructing a model of the MW (including its
luminous components) and then finding the range of values of the local density
that are consistent with all observational data (including, for
instance, rotation curve measurements, velocity dispersions of halo
stars, local surface mass density, total mass)~\cite{co,ggt}. As
emphasised in Ref.~\refcite{ggt} the shape of the MW halo has a crucial
effect on the local density extracted. For a fixed circular speed, in
a flattened halo the local density, in the Galactic plane, is higher.

This work has recently been updated to include new data sets, models
for the MW halo motivated by numerical simulations, and, in some
cases, Bayesian statistical techniques. Ref.~\refcite{wpd}, assuming a
spherical halo with a cuspy density profile ($\rho(r) \propto
r^{-\alpha}$ as $r \rightarrow 0$) found $\rho_{0} = 0.30 \pm 0.05
\, {\rm GeV} \, {\rm cm}^{-3}$, while Ref.~\refcite{cu}, assuming
spherically symmetric
NFW~\cite{nfw} and Einasto~\cite{einasto1,einasto2} profiles for the MW halo
found, $\rho_{0} = 0.39 \pm 0.03 \, {\rm GeV} \, {\rm cm}^{-3}$.
While these determinations have $\sim 10\%$ errors, they
differ from each other by significantly more than this. This suggests
that the systematic errors are bigger than the statistical errors.
Indeed Ref.~\refcite{wdb} finds, considering a range of density profiles including
cored profiles ($\rho(r) \sim {\rm const}$ as $r \rightarrow 0$), 
values in the range  $\rho_{0} = 0.2-0.4 \, {\rm GeV} \, {\rm cm}^{-3}$.

Other recent work has investigated `model independent' techniques,
which don't involve global mass-modelling of the galaxy, and hence
have smaller hidden systematic errors.  Ref.~\refcite{salucci}
proposed using the equation of centrifugal equilibrium and subtracting
the contribution of the stellar component of the MW.  The resulting
determination of the local density is $\rho_{0} = 0.43 \pm 0.11 \pm
0.10 \, {\rm GeV} \, {\rm cm}^{-3}$, where the uncertainties come from
the uncertainty in the slope of the circular speed at the Solar radius
and the ratio of the Solar radius to the length scale of the thin
stellar disk. Ref.~\refcite{garbari} introduced a method which
involves solving the Jeans and Poisson equations with minimal
assumptions. Using Hipparcos data they find $\rho_{0} =
0.11_{-0.27}^{+0.34} \, {\rm GeV} \, {\rm cm}^{-3}$, at $90\%$
confidence, assuming the stellar tracer populations are isothermal and
$\rho_{0} = 1.25_{-0.34}^{+0.30} \, {\rm GeV} \, {\rm cm}^{-3}$ if
they have a non-isothermal profile. An accurate determination of the
vertical dispersion profile of the tracer population is therefore
required to make an unbiased estimate of the local density using this
method.

Ref.~\refcite{pato} used a high-resolution simulation of a Milky Way like
galaxy, including baryons~\cite{patosim}, to investigate the effect of
halo shape on the local density. Specifically they examined how the
local density varies from the density averaged in a spherical shell,
as determined by observations. They find that the density within the
stellar disk, at a distance $R_{0} = 8 \, {\rm kpc}$ from the centre,
is a factor of between 1.01 and 1.41, with an average of 1.21, larger
than the value averaged in a spherical shell.

\subsection{Circular speed}
\label{vc}

The local circular speed, $v_{\rm c} \equiv v_{\rm c}(R_{0})$, is an important quantity to determine. It appears in
the Galilean transformation of the velocity distribution into the lab
frame (see Sec.~\ref{signals}). It is also related
to the radial component of the
velocity dispersion, $\sigma_{r}$, by one of the Jeans equations
(e.g. Ref.~\refcite{bt}):
\begin{equation}
\label{jeans}
 \frac{1}{\rho} \frac{{\rm d} (\rho \sigma_{r}^2)}{{\rm d} r} + 2 \frac{\beta \sigma_{r}^2}{r}  =- \frac{v_{\rm c}^2}{r} \,,
\end{equation}
where the anisotropy parameter $\beta$ is defined in eq.~(\ref{beta}). The SHM has
$\rho(r) \propto r^{-2}$ and the the velocity distribution is isotropic
($\sigma_{r}=\sigma_{\theta}=\sigma_{\phi}$ so that $\beta=0$) and
independent of radius, so that $\sigma_{\rm r}=v_{\rm
  c}/\sqrt{2}$.

The standard value of $v_{\rm c}$ of $220 \, {\rm km \,
  s}^{-1}$ dates back to a 1980s review of the Galactic
constants~\cite{kerr} and was found by taking an average of the values
found from a wide range of different analyses.
Note that the ratio $v_{\rm c}/R_{0}$ is better
determined than either $v_{\rm c}$ or $R_{0}$ individually
(e.g. Refs.~\refcite{reid,mcmillan}).

A recent analysis using measurements of the motions of Galactic
masers, found a significantly higher value, $v_{\rm c} = (254 \pm 16)
\, {\rm km \, s}^{-1}$,~\cite{reid}. Ref.~\refcite{bovy} reanalyzed
the data using a more general model for the maser velocity
distribution (including allowing a non-zero velocity dispersion
tensor) and found that the maser data places only a relatively weak constraint on
$v_{\rm c}$.  When combined with other measurements (from the proper
motion of Sgr ${\rm A}^{\star}$, and the orbit of the GD-1 stellar
stream), they found $v_{\rm c} = (236 \pm 11) {\rm km \, s}^{-1}$,
assuming a flat rotation curve. Meanwhile Ref.~\refcite{mcmillan} found the value of $v_{\rm c}$
determined from the maser data depends strongly on the MW model used.
Using a range of models for the rotation curve, including a power-law
with free slope, they found values in the range
$v_{\rm c} = (200 \pm 20) {\rm km \, s}^{-1}$ to $v_{\rm c} = (279 \pm
33) {\rm km \, s}^{-1}$. This illustrates that, as in the case of the local
density, systematic, modelling errors are important.

\subsection{Escape speed}

The escape speed is the speed required to escape the local
gravitational field of the MW, $v_{\rm esc}(r) = \sqrt{2
  |\Phi(r)|}$. The local escape speed,  $v_{\rm esc} \equiv v_{\rm esc}(R_{0})$, is estimated from the speeds of high velocity stars.
To do this it is necessary to parameterise the shape of the high speed
tail of the velocity distribution. Assuming that the velocities are
isotropic and the Jeans theorem applies, the asymptotic form of the
velocity distribution can be written as~\cite{lt,kochanek}:
\begin{equation}
 f(|{\bf v}|) \propto
\begin{cases}
 (v_{\rm esc}^2 - |{\bf v}|^2)^{k} = [(v_{\rm esc}
- |{\bf v}| ) (v_{\rm esc}+ |{\bf v}|)]^{k}  \,, & |{\bf v}|< v_{\rm
  esc} \,,
\\
 0  \,, &  |{\bf v}| \geq  v_{\rm esc} \,.
\end{cases}
\end{equation}
Traditionally a value $v_{\rm esc} =650 \, {\rm km \, s}^{-1}$,
corresponding to the upper $90\%$ confidence limit from
Ref.~\refcite{lt}, has been used.

Ref.~\refcite{smith} has updated these measurements, using additional
data from the RAVE survey and using a prior on $k$, $k \in [2.7, 4.7]$
(motivated by analysis of the speed distributions of stellar particles
in simulated halos). They find that the escape speed lies in the range
$498 \, {\rm km \, s}^{-1} < v_{\rm esc} < 608 \, {\rm km \, s}^{-1}$
at $90\%$ confidence, with a median likelihood of $v_{\rm esc} =544 \,
{\rm km \, s}^{-1}$.

\section{Simulations}
\label{astro2}

A number of high resolution, dark matter only, simulations of the
formation of Milky Way-like halos in a cosmological context have been
carried out (e.g. Aquarius~\cite{aquarius}, GHALO~\cite{ghalo} and Via
Lactea~\cite{vl}).  The velocity distributions of these halos deviate
systematically from a multivariate
Gaussian~\cite{hansen,fairs,vogelsberger,kuhlen}.  There are more low
speed particles and the peak in the
distribution is lower and flatter in shape (i.e. the distribution is platykurtic)~\footnote{Note,
  however, that the deviation is smaller in the lab frame than in the
  Milky Way rest frame~\cite{kuhlen}.}.  Several fitting functions
have been considered; a Tsallis distribution (which arises from
non-extensive statistical mechanics)~\cite{tsallis,hansen1} and a
modified Maxwellian~\cite{fairs,kuhlen}
\begin{eqnarray}
\label{fvsim}
f(v_{\rm r}) &=& \frac{1}{N_{\rm r}} \exp{ \left[ - \left( \frac{v_{\rm r}^2}{\bar{v}_{\rm r}^2} \right)^{\alpha_{\rm r}} \right]} \,, \\
f(v_{\rm t}) &=& \frac{v_{\rm t}}{N_{\rm t}} \exp{ \left[ - \left( \frac{v_{\rm t}^2}{\bar{v}_{\rm t}^2} \right)^{\alpha_{\rm t}} \right]} \,,
\end{eqnarray}
where $v_{\rm r}$ and $v_{\rm t}=\sqrt{v_{\theta}^2 + v_{\phi}^2}$ are
the radial and tangential components of the velocity, $N_{\rm r/t}$
are normalisation factors and $\alpha_{r,t}$ are free-parameters which
parameterise the deviations from a Maxwellian distribution.  The most
likely speed deviates from the circular speed: for VL2 and GHALO
$v_{\rm c}/v_{0} \approx 0.85$ and $0.86$ respectively~\cite{kuhlen}.

The velocity distributions also have stochastic features at high
speeds. There are broad bumps which vary from halo to halo, but are
independent of position within a given halo and are thought to reflect
the formation history of the halo~\cite{vogelsberger,kuhlen}. Kuhlen
et al.~\cite{kuhlen} also find narrow spikes in some locations,
corresponding to tidal streams.

Ref.~\refcite{lsww} presented an ansatz for the velocity distribution
\begin{equation}
\label{k}
f(|{\bf v}|) \propto \left[ \exp{ \left( \frac{v_{\rm esc}^2 - |{\bf v}|^2}{k
        v_{0}^{2}} \right)} - 1 \right]^{k} \Theta( v_{\rm esc} -
|{\bf v}| ) \,,
\end{equation}
which fits the high speed tail of the smooth component of the
speed distributions found in simulations.
The shape of the high speed tail of the distribution is determined by the slope of
the density profile at large radii. Using the Eddington equation the
parameter $k$ can be related to the outer slope of the density profile,
$\gamma$, ($\rho(r) \propto r^{-\gamma}$ for large $r$), by
$k= \gamma- 3/2$ for $\gamma > 3$~\cite{kochanek}.  As $\gamma \rightarrow 3$ (the value corresponding
to the NFW profile~\cite{nfw}) the calculation breaks down, and numerical fits to eq.~(\ref{k}) find
$k \approx 2$. More generally for outer slopes in the range $\gamma
\sim 3-5$, $k$ lies in the range $k=[1.5-3.5]$. Eq.~(\ref{k}) with $k$
in this range provides a good fit to the high speed tails of the speed
distributions found in simulation and has less high speed particles
than the tail of the standard Maxwellian distribution with a sharp
cut-off. Note,
however, that eq.~(\ref{k}) does not match the low and moderate speed
behaviour of the simulation speed distributions, possibly due to the
assumption of isotropy~\cite{lsww}.

\subsection{Effects of baryons}
\label{baryon}

The simulations discussed above contain dark matter only, while
baryons dominate in the inner regions of the Milky Way. Simulating
baryonic physics is extremely difficult, and producing galaxies whose
detailed properties match those of real galaxies is a major
challenge. Some recent simulations have found that late merging
sub-halos are preferentially dragged towards the disc by dynamical friction, where they are
destroyed leading to the formation of a co-rotating dark disc
(DD)~\cite{read1,read2,ling1}. Ref.~\refcite{bruch} modelled the DD velocity
distribution as a gaussian with isotropic dispersion, $\sigma_{\rm
  DD}=50 \, {\rm km \, s}^{-1}$ and lag $v_{\rm lag} = 50\, {\rm km \,
  s}^{-1}$, matching (roughly) the kinematics of the Milky Way's
stellar thick disc and considered DD densities in the range $0.5
< (\rho_{\rm DD} / \rho_{\rm H}) < 2$, where $\rho_{\rm H}$ is the
local halo density.

The properties (and even existence) of the dark disc are highly
uncertain. Ref.~\refcite{purcell} argues that to be consistent with
the observed morphological and kinematic properties of the Milky Way's
thick disc, the Milky Way's merger history must be quiescent compared
with typical $\Lambda$CDM merger histories, and hence the DD density
must be relatively small, $\rho_{\rm DD} < 0.2 \rho_{\rm H}$.  The
total (halo + disc) local density can be probed by the kinematics of
stars (e.g. Ref.~\refcite{hf}). There is no evidence for a dark matter
disc (i.e. the data are well fit without a dark disc) and a thick,
dense, $(\rho_{\rm DD} / \rho_{\rm H}) > 0.5$, dark disc is
excluded~\cite{bidin}.  Refs.~\refcite{purcell} and \refcite{ling2} also argue that
the DD velocity dispersion is likely to be substantially larger than
that of the stellar thick disc.

Baryonic physics will also affect the halo speed distribution.
For instance gas cooling makes halos more spherical (e.g. Ref.~\refcite{sphere}).
The local speed distribution found in Ref.~\refcite{ling1} is well fit by
a Tsallis distribution, but appears to deviate less from the standard
Maxwellian than the distributions found in dark matter only
simulations.

\subsection{Ultra-local structure}
\label{micro}

A further caution is that the scales resolved by simulations are many
orders of magnitude larger than those relevant for direct detection
experiments.  The Earth moves at $\sim 200 \, {\rm km \, s}^{-1} \sim
0.1 \, {\rm mpc \, yr}^{-1}$, therefore direct detection experiments
probe the dark matter on sub mpc scales, while the numerical
simulations discussed above have gravitational softening of order $10
\, {\rm pc}$.

Vogelsberger and White have developed a new technique to follow the
fine-grained phase space distribution in simulations as it stretches
and folds under the action of gravity~\cite{vw}. They find that the
median density of the resulting streams is of order $10^{-7}$ the
local halo density. Schneider et al.~\cite{skm} have studied the
evolution of the first, and smallest, roughly Earth mass~\cite{hss,ghs1,ghs2,dms},
microhalos to form in the Universe.  They find that tidal disruption
and encounters with stars produce tidal streams with average density
$\sim 10^{-4}$ the local halo density. These results (see also
Ref.~\refcite{kk2}) suggest that the ultra-local dark matter density
and velocity distribution should not be drastically different to those
on the scales resolved by simulations. The ultra-local velocity
distribution may, however, contain some features or fine-grained
structure~\cite{fantin1,afshordi1,afshordi2,lissperg,fantin2}.

\section{Differential event rate and signals}
\label{signals}

The differential event rate for elastic scattering, assuming
spin-independent coupling with identical couplings to the proton and
neutron, is given by (e.g. Refs.~\refcite{jkg,ls}):
\begin{equation}
\label{drde}
\frac{{\rm d} R}{{\rm d}E}(E,t) =
             \frac{\sigma_{{\rm p}} 
             \rho_{0}}{2 \mu_{{\rm p} \chi}^2 m_{\chi}}
             A^2 F^2(E)   \int_{|{\bf v}| \geq v_{{\rm min}}} 
            \frac{f({\bf v},t)}{v} {\rm d}^3 {\bf v}     \,, 
\end{equation}
where $\rho_{0}$ is the local WIMP density, $f({\bf v}, t)$ the 
WIMP velocity distribution in the lab frame,
$\sigma_{{\rm p}}$ the WIMP scattering cross section on the proton,
$\mu_{{\rm p} \chi} = (m_{\rm p} m_{\chi})/(m_{{\rm p}}+ m_{{\chi}})$
the WIMP-proton reduced mass, $A$ and $F(E)$ the mass number and form
factor of the target nuclei respectively and $E$ is the recoil energy.
The lower limit of the integral, $v_{{\rm min}}$, is the minimum WIMP
speed that can cause a recoil of energy $E$:
\begin{equation}
\label{vmin}
v_{{\rm min}}=\left( \frac{ E m_{A}}{2 \mu_{{\rm A} \chi}^2} 
             \right)^{1/2} \,,
\end{equation}
where $m_{A}$ is the atomic mass of the detector nuclei
and $\mu_{{\rm A} \chi}$ the WIMP-nucleon reduced mass. The event rate
for inelastic scattering of WIMPs~\cite{inelastic} also depends on the
WIMP density and velocity distribution. However, due to the altered
kinematics the relationship between $v_{\rm min}$ and $E$ is
changed, and hence the effects of changes in the velocity distribution are different~\cite{mmm,kuhlen,ling2,mccabe}.

The WIMP speed distribution must be transformed from the Galactic rest
frame to the lab frame. This is done by carrying out, a time
dependent, Galilean transformation: ${\bf v} \rightarrow
\tilde{\bf{v}} = {\bf v} + {\bf v}_{e}(t)$~\footnote{Formally gravitational focusing by the Sun should be taken into
  account~\cite{griest,ag2}, however the resulting modulation in the
  differential event rate is small and only detectable with a very
  large number of events~\cite{griest}.}.  The Earth's motion relative
to the Galactic rest frame, ${\bf v}_{e}(t)$, is made up of three
components: the motion of the Local Standard of Rest (LSR), ${\bf
  v}_{\rm LSR}$, the Sun's peculiar motion with respect to the LSR,
${\bf v}_{\odot}^{\rm p}$, and the Earth's orbit about the Sun, ${\bf
  v}_{e}^{\rm orb}$,.  The motion of the LSR is defined as ${\bf
  v}_{\rm LSR}=(0, v_{\rm c}, 0)$, where $v_{\rm c}$ is the local
circular speed (see Sec.~\ref{vc} for a discussion of recent
determinations).  The most recent determination of the Sun's motion with
respect to the LSR, taking into account the effects of the metallicity
gradient in the disc, finds $ {\bf v}_{\odot}^{\rm p} = (U_{\odot},
V_{\odot}, W_{\odot}) = (11.1, \, 12.2, 7.3) \, {\rm km \,
  s}^{-1}$~\cite{schoenrich} in Galactic co-ordinates (where $U$
points towards the Galactic center, $V$ is the direction of Galactic
rotation and $W$ towards the North Galactic Pole). Note that the value
of $V_{\odot}$ is significantly ($\sim 7 \, {\rm km \, s}^{-1}$)
larger than previously found~\cite{db}. The new larger value is also
supported by the analysis of the motions of masers~\cite{mcmillan}.
Accurate expressions for the Earth's orbit can be found in
Ref.~\refcite{book}.  Simpler expressions, which are acceptable for most
practical purposes, can be found in Refs.~\refcite{fs,gg}.

The differential event rate, eq.~(\ref{drde}), depends on the target
nuclei mass, the (a priori unknown) WIMP mass and the integral of the
velocity distribution. It is therefore useful to rewrite eq.~(\ref{drde}) as 
\begin{equation}
\frac{{\rm d} R}{{\rm d}E}(E,t) = C_{{\chi}, {\rm A}} \,
  \rho_{0} \,  g(v_{\rm min},t ) \,,
\end{equation}
where
\begin{equation}
\label{gvmin}
g(v_{\rm min},t) =  \int_{|{\bf v}| \geq v_{\rm min}} \frac{f({\bf v},t)}{v}
\, {\rm d}^{3} {\bf v} \,,
\end{equation}
and  the prefactor
\begin{equation}
\label{C}
C_{{\chi}, {\rm A}}= \frac{\sigma_{{\rm p}}}{2 \mu_{{\rm p} \chi}^2 m_{\chi}}
             A^2 F^2(E) \,,
\end{equation}
contains the WIMP and target dependent terms and is independent of the
astrophysical WIMP distribution.

We will now discuss the energy (Sec.~\ref{er}), time (Sec.~\ref{am})  and 
direction (Sec.~\ref{dd}) dependence of the differential event rate. In order
to make concrete statements we will for now assume the standard Maxwellian
velocity distribution, eq.~(\ref{max}). In Sec.~\ref{affect} we will
discuss how uncertainties in the WIMP distribution affect these signals.

\subsection{Energy dependence}
\label{er}

The shape of the energy spectrum depends on both the WIMP mass and the
mass of the target nuclei. This can be seen, following Lewin and
Smith~\cite{ls}, by assuming a standard Maxwellian velocity
distribution, eq.~(\ref{max}), and, initially, neglecting the Earth's
velocity and the Galactic escape speed. This allows the time averaged
energy spectrum
to be written as
\begin{equation}
\label{drde0}
\frac{{\rm d} R}{{\rm d}E}(E) = \left(\frac{{\rm d} R}{{\rm d}E}\right)_{0}
              \exp{ \left( -\frac{E}{E_{\rm R}} \right)}
              F^2(E)  \,,  
\end{equation}
where $({\rm d} R/{\rm d}E)_{0}$ is the event
rate in the $E \rightarrow 0\, {\rm keV
}$ limit, and $E_{\rm R}$, the characteristic
energy scale, is given by
\begin{equation}
\label{ereq}
E_{\rm R} = \frac{2 \mu_{{\rm A} \chi}^2 v_{c}^2}{m_{\rm A}} \,.
\end{equation}
When the Earth's velocity and the Galactic escape speed are taken into
account eq.~(\ref{drde0}) is still a reasonable approximation to the
event rate if $({\rm d} R/ {\rm d} E )_{0}$ and
$E_{\rm R}$ are both multiplied by constants of order unity.  If $m_{\chi}
\ll m_{\rm A}$, $E_{\rm R} \propto m_{\chi}^{2}$, while if $m_{\chi}
\gg m_{\rm A}$, $E_{\rm R}$ is independent of the WIMP mass. This
indicates that the WIMP mass can be determined from the energy spectrum,
provided it is not significantly larger than the mass of the target
nuclei~\cite{wimpmass1,wimpmassme,dreesshan}. Furthermore measuring consistent spectra 
for two different target nuclei could in principle confirm the WIMP origin of
these spectra (e.g. Ref.~\refcite{matsig}). This is sometimes referred to as the
`materials signal'.

\subsection{Annual modulation}
\label{am}

Due to the Earth's orbit about the Sun, the net velocity of the lab
with respect to the Galactic rest frame varies annually. The net speed is largest in the Summer and hence there are more
high speed WIMPs, and less low speed WIMPs, in the lab frame. This
produces an annual modulation of the event rate~\cite{dfs,ffg}.  The
differential event rate peaks in Winter for small recoil energies and
in Summer for larger recoil energies~\cite{primack}. The energy
at which the annual modulation changes phase is often referred to as
the `crossing energy'. Its value depends on the WIMP and target
masses, and hence could be used to determine the WIMP mass~\cite{lewis}.

Since the Earth's orbital speed is significantly
smaller than the Sun's circular speed the amplitude of the modulation
is small and, to a first approximation, the differential event rate
can, for the standard halo model, be written approximately as a Taylor
series:
\begin{equation}
\frac{{\rm d} R}{{\rm d} E}(E,t) \approx 
   \frac{{\rm d} R}{{\rm d} E} (E)
  \left[ 1 + \Delta(E) 
  \cos{\alpha(t)} \right] \,,
\end{equation}
where $\alpha(t) = 2 \pi (t-t_{0})/T$, $T=1$ year and $t_{0} \sim 150$
days.
For the standard halo model the fractional amplitude of the modulation is
approximately given by~\cite{savage} 
\begin{equation}
\Delta(E)  \approx
\begin{cases}
-0.034 \left( 1 - \frac{x^2}{x^2_{\rm p}} \right) & x< x_{\rm p}
\nonumber \\
0.014 \left( \frac{x}{x_{\rm p}} -1 \right) \left(   \frac{x}{x_{\rm
      p}} +3.7 \right) & x_{\rm p} < x \lesssim z 
\end{cases}
\end{equation}
where $x = v_{\rm min}/v_{\rm c} $, $x_{\rm p}= 0.89$ is the value of $x$ at which the sign
of the modulation reverses and $z= v_{\rm esc} /v_{\rm c} $. For
$v_{\rm min} \gtrsim v_{\rm esc}$, in the extreme tail of the speed
distribution, the shape of the modulation is non-sinusoidal.

For very small energies $\Delta(E)$ is negative (i.e. the maximum
occurs in December rather than June). As $E$ is increased,
$|\Delta(E)|$ initially decreases to zero at which point the phase of
the annual modulation changes so that the maximum occurs in Summer. As
$E$ is increased further the fractional amplitude continues
increasing. This is potentially misleading however, as the mean event
rate, and hence the raw amplitude, becomes very small at large $E$.
After the phase change on increasing $E$ further the raw amplitude
increases to a local maximum before decreasing again and tending to
zero. For the SHM, for measurable energies, the amplitude of the modulation lies in
the range $1-10 \%$.

\subsection{Direction dependence}
\label{dd}

Our motion with respect to the Galactic rest frame also produces a
direction dependence of the event rate.  The WIMP flux in the lab
frame is peaked in the direction of motion of the Sun (towards
the constellation CYGNUS), and hence the recoil spectrum is, for most
energies~\footnote{At low energies the maximum recoil rate is in a
  ring around the average WIMP arrival direction~\cite{bgg1,bgg2}.}, peaked in
the direction opposite to this~\cite{spergel}. This directional signal
is far larger than the annual modulation; the event rate in the
backward direction is several times larger than that in the forward
direction~\cite{spergel}. A detector which can measure the recoil
directions is required to detect this signal (see Ref.~\refcite{ahlen}
for an overview of the status of directional detection experiments).

The full direction dependence of the event rate is most compactly
written in terms of the radon transform of the WIMP velocity
distribution~\cite{gondolo}
\begin{equation}
\frac{{\rm d}R}{{\rm d} E \, {\rm d} \Omega} = 
\frac{\rho_{0} \sigma_{\rm p} A^2}{4 \pi \mu_{{\rm p} \chi}^2 m_{\chi}} F^2(E) 
   \hat{f}(v_{\rm min},\hat{{\bf q}}) \,,
\end{equation}
where ${\rm d} \Omega = {\rm d} \phi \, {\rm d} \cos{\gamma}$,
$\hat{\bf{q}}$ is the recoil direction and
$\hat{f}(v_{\rm min},\hat{{\bf q}})$ is the 3-dimensional Radon transform
of the WIMP velocity distribution 
\begin{equation}
\hat{f}(v_{\rm min},\hat{{\bf q}}) = \int \delta({\bf v}.\hat{{\bf q}} - v_{\rm min}) f({\bf v})
 {\rm d}^3 {\bf v} \,.
\end{equation}
Geometrically the Radon transform, $\hat{f}(v_{\rm min},\hat{{\bf
    q}})$, is the integral of the function $f({\bf v})$ on a plane
orthogonal to the direction $\hat{\bf{q}}$ at a distance $v_{\rm min}$
from the origin. See Ref.~\refcite{ck2} for an alternative, but
equivalent, expression.

While the directional recoil rate depends on both of the angles which
specify a given direction, the strongest signal is the differential of
the event rate with
respect to the angle between the recoil and the direction of solar
motion, $\gamma$,~\cite{spergel,ag}
\begin{equation}
\label{drdedcos}
\frac{{\rm d}^2 R}{{\rm d} E \, {\rm d}\cos{\gamma}} 
  = \frac{\rho_{0} \sigma_{\rm p}}{4 \pi \mu_{{\rm p}\chi} m_{\chi}} A^2 F^2(E) 
  \int_{0}^{2 \pi}   \hat{f}(v_{\rm min},\hat{{\bf q}}) \, {\rm d} \phi \,.
\end{equation}
For the standard Maxwellian velocity distribution~\cite{spergel,ag}
\begin{equation}
\frac{{\rm d}^2 R}{{\rm d} E \, {\rm d}\cos{\gamma}}  = 
\frac{C(\chi, A) \rho_{0}}{\sqrt{\pi} v_{\rm c}} \exp{ \left[ -
    \left( \frac{ (v_{\rm e}^{\rm orb, p}  + v_{\rm c}) \cos{\gamma} -
        v_{\rm min} }{v_{\rm c}} \right)^2 \right]} \,, 
\end{equation}
where $v_{\rm e}^{\rm orb, p}$ is the component of the Earth's
velocity parallel to the direction of Solar motion.

An ideal detector capable of measuring the nuclear recoil vectors
(including their senses, $+{\bf q}$ versus $-{\bf q}$) in 3-dimensions,
with good angular resolution, could reject isotropy of WIMP recoils
with only of order 10 events~\cite{copi:krauss,ck2,pap1}.  Most, but not
all, backgrounds would produce an isotropic Galactic recoil
distribution (due to the complicated motion of the Earth with respect
to the Galactic rest frame).  An anisotropic Galactic recoil
distribution would therefore provide strong, but not conclusive,
evidence for a Galactic origin of the recoils. Roughly 30 events would
be required for an ideal detector to confirm that the peak recoil
direction coincides with the inverse of the direction of Solar motion, hence
confirming the Galactic origin of the recoil events~\cite{billard,pap5}.

\section{Consequences}
\label{affect}

In this section we will discuss how the uncertainties described in
Secs.~\ref{astro1} and \ref{astro2} affect the signals expected in
direct detection experiments.

\subsection{Energy dependence}

Since the normalisation of the event rate is directly proportional to
the product of the cross-section, $\sigma_{\rm p}$, and the local
density, $\rho_{0}$, the uncertainty in $\rho_{0}$ leads directly to an
uncertainty in measurements of, or constraints on, $\sigma_{\rm
  p}$. Since the density is not expected to vary on mpc scales,
the uncertainty in $\sigma_{\rm p}$ is the same for all direct
detection experiments,
and does not affect comparisons of e.g. the exclusion limits from
different experiments. It does however affect comparisons with
collider and other constraints on $\sigma_{\rm p}$. Furthermore it can
significantly bias the determination of the WIMP mass~\cite{st}.

The characteristic energy scale, $E_{\rm R}$, given by eq.~(\ref{ereq}),
depends on the local circular speed. Therefore the uncertainty in the
local circular speed, $v_{\rm c}$, leads to a ${\cal O}(10 \%)$
uncertainty in the differential event rate, and hence exclusion
limits~\cite{mccabe,greenfv}. The nature of the change in the
exclusion limits is similar, but not identical, for different
experiments~\cite{greenexclude,mccabe}. It will also lead to a bias in 
determinations of the WIMP mass~\cite{wimpmass1,wimpmassme}. This can
be seen by differentiating the expression for the characteristic
energy, eq.~(\ref{ereq}),
\begin{equation}
\frac{\Delta m_{\chi}}{m_{\chi}} \sim - \left[ 1 + \left(\frac{m_{\chi}}{m_{\rm
    A}}\right)\right] \frac{\Delta v_{\rm c}}{v_{\rm c}} \,.
 \end{equation}

 Since the differential event rate is given by an integral over the
 velocity distribution the differential event rate depends only weakly on
 the detailed form of the velocity
 distribution~\cite{kk,donato}. Consequently the resulting uncertainty in
 exclusion limits~\cite{greenexclude,mccabe} and determinations of
 the WIMP mass~\cite{wimpmassme} is usually fairly small. There are,
 however, some exceptions to this statement. Firstly if the WIMP is
 light and/or the experimental energy threshold is high compared with
 the characteristic energy, then only the high speed tail of the speed
 distribution is probed. In this case the uncertainties in its shape
 can have a significant effect on the expected energy spectrum and
 hence exclusions limits or allowed regions~\cite{mccabe,lsww}.
 Secondly if there is a dark disc there will be an additional population of
 low speed WIMPs. If the dark disc density is sufficiently high and
 its velocity dispersion sufficiently small this will significantly
 change the energy spectrum and hence exclusion limits and mass
 determinations~\cite{greenfv}. The size of these changes depends on
 the properties of the dark disc, which are currently
 uncertain~\cite{purcell,ling2}. A WIMP stream, with sufficiently high
 density, would add a sloping step to the differential event rate~\cite{gg}.

\subsection{Annual modulation}

The annual modulation arises from the, relatively small, shift in the
speed distribution in the lab frame over the course of the year. It is
therefore far more sensitive to the speed distribution than the time
averaged energy spectrum. Both the amplitude and phase of the
modulation, and hence the regions of parameter space compatible with
an observed signal, can vary significantly
(e.g. Refs.~\refcite{br,vergados1,belli1,greenam1,vergados2,greenam2,belli2,vergados3,fs,greenfv,cbc}).

The uncertainty in the value of $v_{\rm c} $ can change the
amplitude by a factor of order unity, while the uncertainty in the
shape of the halo velocity distribution changes the amplitude by
$\sim {\cal O}(10 \%)$~\cite{greenam1,greenfv}.  A high density dark
disc could change the annual modulation signal significantly,
producing a 2nd, low energy, maximum in the amplitude of the
modulation and changing the phase significantly~\cite{greenfv}.
For a WIMP stream the position
and height of the step in the energy spectrum produced would vary
annually (e.g. Ref.~\refcite{savage}).

\subsection{Direction dependence}

The detailed direction dependence of the event rate is sensitive to
the velocity distribution, however the anisotropy is
robust~\cite{copi:krauss,ck2,pap1}. The number of events required to
detect anisotropy or to demonstrate that the median inverse recoil
direction coincides with the direction of solar motion only vary by of
order $10\%$~\cite{copi:krauss,ck2,pap1,pap5}.

The features in the high speed tail can cause the median inverse
direction of high energy recoils to deviate from the direction of
solar motion~\cite{kuhlen}. A stream of WIMPs would produce recoils
which are strongly peaked in the opposite direction~\cite{ag2}. In the
long term, studying the median direction of high energy recoils could
allow high speed features to be detected and the formation
history of the Milky Way probed.

\section{How to handle the uncertainties?}
\label{strategies}

\subsection{Astrophysics independent methods}

It is possible to carry out an `astrophysics independent' comparison
of data from multiple experiments. This effectively involves making
comparisons of the energy spectra, so that the integral over the velocity
distribution, $g(v_{\rm min})$ defined in eq.~(\ref{gvmin}), cancels.
For instance, with data from two different experiments using two
different targets, the WIMP mass can in principle be determined,
without any assumptions about $f({\bf v})$, by taking moments of the energy
spectra~\cite{dreesshan}. However this method leads to a systematic
underestimate of the WIMP mass if it is comparable to or larger than the mass of the
target nuclei.

Ref.~\refcite{flw}, see also Ref.~\refcite{fkt}, showed how the differential event rates in two
experiments, `1' and `2' with target nuclei with mass numbers $A_{1}$ and
$A_{2}$ are related. If the first experiment is sensitive to recoil
energies in the range $[E_{1}^{\rm low}, E_{1}^{\rm high}]$ it probes
$g(v_{\rm min})$ for $v_{{\rm min}}^{\rm low} \leq v_{\rm min} \leq v_{{\rm
    min}}^{\rm high}$, where $v_{\rm min}$ is related to the recoil energy,
WIMP mass and target nuclei mass by eq.~(\ref{vmin}). In experiment 2
these values of $v_{\rm min}$ correspond to recoils in the range
\begin{equation}
[E_{2}^{\rm low}, E_{2}^{\rm high} ] = \frac{\mu_{A2 \chi}^2
        m_{A1}}{\mu_{A1 \chi}^2 m_{A2}} [E_{1}^{\rm low} , E_{1}^{\rm
        high}] \,.
\end{equation}
For recoil energies in this range the
differential event rates in the two experiments are then related by
\begin{equation}
\frac{{\rm d} R_{2}}{{\rm d} E}(E_{2}) = \frac{C_{\chi, A_{2}}}{C_{\chi,
      A_{1}}} \frac{ F_{2}^2(E_{2})}{F_{1}^2 \left( \frac{ \mu_{A1 \chi}^2
        m_{A2}}{\mu_{A2 \chi}^2 m_{A1}} E_{2} \right)} \frac{{\rm d}
      R_{1}}{{\rm d} E}\left( \frac{ \mu_{A1 \chi}^2
        m_{A2}}{\mu_{A2 \chi}^2 m_{A1}}  E_{2} \right) \,.
\end{equation}
This relation depends on the (unknown) WIMP mass, therefore the
comparison has to be made separately for a range of $m_{\chi}$ values.
This method has recently been extended to include annual modulation data~\cite{oxam}.

This approach has proved particularly useful~\cite{fklw,mccabe2,oxam} in
comparing the event rate excesses and annual modulations found in the
CoGeNT~\cite{cogent1,cogent2}, CRESST~\cite{cresst} and DAMA~\cite{dama}
experiments with exclusions limits from CDMS~\cite{cdms} and
Xenon~\cite{xenon1,xenon2}.  Ref.~\refcite{oxam} found that the DAMA, CoGeNT
and CRESST-II data are compatible which each other, but not CDMS and
XENON, if the velocity distribution is very anisotropic (which leads
to a large modulation fraction).

\subsection{Parameterizing the speed distribution}

The astrophysics independent methods are invaluable for assessing the
compatibility of data from different experiments. However the WIMP
mass is not a priori known, and the goal of WIMP direct detection
experiments is not just to detect WIMPs, but to also measure their
mass and cross-section. While there are uncertainties in the velocity
distribution it is not completely unknown, and therefore does not need
to be completely removed from the analysis.

Strigari and Trotta~\cite{st} have shown how the WIMP properties can
be constrained by a single future direct detection experiment when
astronomical data (such as the kinematics of halo stars) are used to
jointly constrain the WIMP properties and the parameters of a model
for the Milky Way halo. They assumed an isotropic Maxwellian speed
distribution characterised by the peak speed, $v_{0}$, and the escape
speed, $v_{\rm esc}$.

Peter~\cite{peter1,peter2} has examined how data sets from multiple
direct detection experiments could be used to jointly constrain a
Maxwellian parametrisation of the WIMP speed distribution and the WIMP
parameters (mass and cross-section). Pato et al.~\cite{pato2} have
taken a similar approach using the parameterization in
eq.~(\ref{k}). Peter has recently extended this work by allowing the
peak speed and circular speed to differ (as expected if the density
profile is not $\rho(r) \propto r^{-2}$) and also considering an
empirical speed distribution consisting of a five or ten bin step
function~\cite{peter2}.

Combining data from multiple experiments with different targets
significantly increases the accuracy with which the WIMP parameters
can be measured~\cite{peter1,pato}.  However fixing the form of the
speed distribution leads to biases in the WIMP properties, if the true
speed distribution differs significantly from the parameterization
assumed~\cite{peter2}.  With a suitable parameterization o the
WIMP parameters and speed distribution can be jointly
probed~\cite{peter2}. The form of the optimal parameterization of
the speed distribution is an open question.


\section{Summary}

Direct detection event rate calculations often assume the standard
halo model, with an isotropic Maxwellian velocity distribution with
dispersion $\sigma = \sqrt{3/2} \, v_{\rm c} = 270 \, {\rm km \, s}^{-1}$ and a local WIMP
density $\rho_{0}= 0.3 \, {\rm GeV} \, {\rm cm}^{-3}$. However it has long
been realised~\cite{dfs} that uncertainties in the speed distribution
will affect the signals expected in experiments.

We have discussed the standard halo model and other approaches to halo
modelling, and the assumptions behind them. We then reviewed
observational determinations of quantities that are relevant to direct
detection experiments, namely the local dark matter density, the local
circular speed (which is related to the velocity dispersion by the
Jeans equations) and the local escape speed. While the statistical
errors on these quantities are often small the systematic errors, from
uncertainties in the modelling of the Milky Way, can be significantly
larger.

Next we turned our attention to high resolution, dark matter only
simulations of the formation of Milky Way-like halos in a cosmological
context. They find velocity distributions that deviate
significantly from the standard Maxwellian, and have features at high
speeds.  The effect of baryonic physics on the dark matter
distribution is not yet well understood. Some recent simulations have
found that a dark disc may be formed, however its properties (and even
existence) are highly uncertain. Simulations resolve the dark matter
distribution on scales many orders of magnitude larger than those
probed by direct detection experiments. The latest results suggest
that the ultra-local dark matter distribution is largely smooth, but
some features may exist.

We then reviewed the resulting uncertainties in the energy, time and
direction dependence of the energy spectrum, and hence constraints on,
or measurements of, the WIMP parameters. The uncertainty in the local
circular speed has a significant effect on the event rate and hence
exclusion limits and determinations of the WIMP mass. The uncertainty
in the local density leads directly to an uncertainty in measurements
of, or constraints on, the cross-section. Only the high energy tail of
the energy spectrum is particularly sensitive to the exact shape of
the speed distribution. The annual modulation is far more sensitive to
the shape of the velocity distribution; its amplitude and phase can
change significantly. The anisotropy of the recoil rate is robust to
changes in the velocity distribution, however high speed features can change
the peak direction of high energy recoils, and hence provide a way of
probing the formation of the Milky Way.

Finally we discussed techniques for handling the astrophysical
uncertainties when analysing direct detection data. Astrophysics
independent comparisons between different experiments can be made
using the integral of the velocity distribution, $g(v_{\rm min})$. This approach
is extremely useful for assessing the compatibility of various
experiments, but requires the WIMP mass as input. Parameterising the
WIMP speed distribution, and using data from multiple experiments to
jointly constrain the WIMP mass and cross-section and speed
distribution is a promising approach, but the optimal form for the
parameterisation is not yet known.

\section*{Acknowledgments}

The author is supported by STFC. She is grateful to the organisers of, and
participants in, the `Dark matter underground and in the heavens' workshop (DMUH11) at
CERN, where she had various useful discussions about the topics
covered in this review. She is also grateful to Bradley Kavanagh for
useful comments on a draft of this article.

\end{document}